\documentclass[12pt,preprint]{aastex}

\shorttitle{Quasi-periodic slipping magnetic reconnection}
\shortauthors{Li \& Zhang}

\begin{document}

\title{Quasi-periodic Slipping Magnetic Reconnection During an X-class Solar Flare Observed by the \emph{Solar Dynamics Observatory} and \emph{Interface Region Imaging Spectrograph}}

\author{Ting Li\altaffilmark{} \& Jun Zhang\altaffilmark{}}

\altaffiltext{}{Key Laboratory of Solar Activity, National
Astronomical Observatories, Chinese Academy of Sciences, Beijing
100012, China; [liting;zjun]@nao.cas.cn}

\begin{abstract}

We firstly report the quasi-periodic slipping motion of flare loops
during an eruptive X-class flare on 2014 September 10. The slipping
motion was investigated at a specific location along one of the two
ribbons and can be observed throughout the impulsive phase of the
flare. The apparent slipping velocity was 20$-$110 km s$^{-1}$ and
the associated period was 3$-$6 min. The footpoints of flare loops
appeared as small-scale bright knots observed in 1400 {\AA},
corresponding to fine structures of the flare ribbon. These bright
knots were observed to move along the southern part of the longer
ribbon and also exhibited a quasi-periodic pattern. The Si {\sc iv}
1402.77 {\AA} line was redshifted by 30$-$50 km s$^{-1}$ at the
locations of moving knots with a $\sim$40$-$60 km s$^{-1}$ line
width, larger than other sites of the flare ribbon. We suggest that
the quasi-periodic slipping reconnection is involved in this process
and the redshift at the bright knots is probably indicative of
reconnection downflow. The emission line of Si {\sc iv} at the
northern part of the longer ribbon also exhibited obvious redshifts
of about 10$-$70 km s$^{-1}$ in the impulsive phase of the flare,
with the redshifts at the outer edges of the ribbon larger than
those in the middle. The redshift velocities at post-flare loops
reached about 80$-$100 km s$^{-1}$ in the transition region.

\end{abstract}


\section{Introduction}

Solar flares result from the abrupt release of magnetic energy via
the process of magnetic reconnection (Parker 1957; Zweibel \& Yamada
2009). Eruptive flares are accompanied by coronal mass ejections
(CMEs) that constitute major drivers for space weather (Bothmer \&
Schwenn 1994; Forbes et al. 2006). The classical two-dimensional
(2D) magnetic reconnection model called the CSHKP model (e.g., see
the review of Shibata \& Magara 2011) has been developed to explain
some characteristics of eruptive flares, including the formation of
coronal flux ropes (e.g., Sakurai 1976), flare ribbons (e.g.,
Schmieder et al. 1996) and post-flare loops (e.g., Milligan \&
Dennis 2009). Flare ribbons are correlated with the impact of
energetic particles launched from the reconnection site underlying
the eruptive flux rope with the chromosphere (Asai et al. 2004; Reid
et al. 2012; Zhang et al. 2014). The velocity behavior measured in
chromospheric ribbons is blueshifted surrounded by redshifts due to
the post-flare loops material (Schmieder et al. 1987). After the
reconnection the material is cooling and descending along the loops.
In H$\alpha$ the velocity along the post-flare loops is of the order
of 100$-$150 km s$^{-1}$ (Schmieder et al. 1996; Wiik et al. 1996;
van Driel-Gesztelyi et al. 1997). It is nearly the free fall value.

However, the 2D CSHKP model can not completely explain the intrinsic
three-dimensional (3D) nature and the complex evolution of eruptive
flares, such as the formation of the sigmoid (Green et al. 2011),
double J-shaped ribbons and strong-to-weak shear transition in
post-flare loops (Aulanier et al. 2012). Recently, 3D extensions to
the CSHKP model have been proposed by Aulanier et al. (2012) and
Janvier et al. (2013). They have presented the 3D
magnetohydrodynamic (MHD) evolution process of the flux rope
ejection and the slipping motion of field lines in 3D reconnection.
Magnetic field lines undergo a continuous series of reconnections as
they cross the quasi-separatrix layers (QSLs; Priest \& D{\'e}moulin
1995; D{\'e}moulin et al. 1996), where field line linkage displays a
rapid change but is not necessarily discontinuous as in
separatrices. The continuous restructuring of field lines along the
QSLs results in the apparent slipping motion of field line
footpoints (Aulanier et al. 2005). Depending on the velocity of the
slipping motion, two definitions were given in Aulanier et al.
(2006): if the velocity is sub-Alfv\'{e}nic, this is the so-called
slipping reconnection, while if the velocity is super-Alfv\'{e}nic,
such slipping field lines correspond to the slip-running
reconnection regime.

The observational studies about slipping magnetic reconnection are
very rare until now, although the theoretical models have been well
developed. Aulanier et al. (2007) and Dud{\'{\i}}k et al. (2014)
respectively investigated the apparent slipping motions of bright
coronal loops in active region (AR) and flare loops during an
eruptive flare, supporting the existence of slipping magnetic
reconnection. Based on the \emph{Solar Dynamics Observatory}
(\emph{SDO}; Pesnell et al. 2012) observations, Li \& Zhang (2014)
reported the slipping motion of flare loops during the eruption of a
flux rope. Different from previous studies, the slipping motion of
flare loops in this work shows a quasi-periodic pattern. Here, the
quasi-periodic pattern means that the appearance of flare loops and
their subsequent slipping motion are intermittent. The recently
launched \emph{Interface Region Imaging Spectrograph} (\emph{IRIS};
De Pontieu et al. 2014) mission is now providing observations of the
transition region (TR) and the chromosphere with remarkable spatial
and spectral resolution. Here, we present the first simultaneous
imaging and spectroscopic observations of quasi-periodic slipping
magnetic reconnection based on \emph{SDO} and \emph{IRIS} data.

\section{Observations and Data Analysis}

The Atmospheric Imaging Assembly (AIA; Lemen et al. 2012) onboard
the \emph{SDO} uninterruptedly observes the full disk of the Sun in
10 (E)UV channels at 1$\arcsec$.5 resolution and 12 s cadence. We
use the observations of 131, 94, 171, 193 and 1600 {\AA} on 2014
September 10 to investigate the evolutions of flare loops and
ribbons. The full-disk line-of-sight magnetograms from the
Helioseismic and Magnetic Imager (HMI; Scherrer et al. 2012) are
also applied. The \emph{IRIS} observations were taken from 11:28 UT
to 17:58 UT, with the field of view (FOV) covering the majority of
the NOAA AR 12158. The slit-jaw images (SJIs) centered at 1400 {\AA}
are used to analyze the fine-scale structures of flare ribbons, with
0$\arcsec$.33$-$0$\arcsec$.4 spatial resolution and a cadence of
$\sim$ 19 s (De Pontieu et al. 2014). The spectral data are taken in
a sit-and-stare mode with 9 s cadence and a spectral dispersion of
$\sim$0.026 {\AA} pixel$^{-1}$. The emission line of Si {\sc iv} is
mainly analyzed. The Si {\sc iv} 1402.77 {\AA} line is formed in the
middle TR with a temperature of about 10$^{4.9}$ K (Tian et al.
2014a; Li et al. 2014). Since the profile of Si {\sc iv} 1402.77
{\AA} line is close to Gaussian to some extent, we applied a
single-Gaussian fit to the 1402.77 {\AA} line (Peter et al. 2014)
and obtained the temporal evolution of peak intensity, Doppler
shift, and line width (Figures 4 and 5).

\section{Results}

\subsection{Overview of the X1.6 Flare}

On 2014 September 10, an X1.6 flare occurred in the sigmoidal region
NOAA AR 12158 (Cheng et al. 2015), including an $\eta$-shaped
negative-polarity flare ribbon (north part as NNR and south part SNR
in Figures 1(g)-(h)) and a semicircular positive-polarity ribbon (PR
in Figures 1(g)-(h)) surrounding the sunspot. The GOES soft X-ray
1$-$8 {\AA} flux showed that the X1.6 flare initiated at 17:21 UT,
reached its peak at 17:45 UT (Figure 3(d)). Two groups of loop
systems (``G1" and ``G2" in Figure 1; see Animation 131-slippage)
observed at 131 {\AA} and 94 {\AA} were involved in the flare.
During the flare evolution, more flare loops successively appeared
and the two loop systems ``G1" and ``G2" both exhibited apparent
slipping motions. The south footpoints of ``G1" gradually propagated
towards the southwest (panels (a)-(f)), which develops into the
ribbon SNR (panels (h)-(i)). Similarly, the south footpoints of
``G2" slipped to the southwest (panels (a)-(f)) and corresponded to
the propagation of the NNR (panels (h)-(i)). The flare loops
generally refer to the hot coronal loops that are formed between the
two ribbons, the counterpart of the loops that belong to the CME and
that stay in the lower atmoshpere (they used to be call post-flare
loops). However, the flare loops in this work are moving to the tip
of the ribbons and these are not post-flare loops.

\subsection{Quasi-periodic Slipping Motion}

Immediately at the start of the flare, the flare loops of the system
``G1" exhibited the apparent slipping motion (Figures 2(a)-(b)), and
their footpoints developed into the SNR (panels (c)-(d)). The
fine-scale structures of the SNR could be observed in \emph{IRIS}
1400 {\AA} images, appearing as numerous dot-like features (panels
(e)-(h); see Animation 1400-ribbon). These small-scale knots were
seen to appear intermittently and propagate along the ribbon towards
the southwest (diamond and circle symbols in panels (e)-(h)). In
order to analyze the kinematic evolution of the slipping loops in
AIA EUV channels in detail, we obtain the stack plots in different
wavelengths (Figures 3(a)-(c)) along the straight line ``A$-$B"
(Figure 1(f)). The stack plots show multiple moving intensity
features, with each strip representing the apparent slippage of
flare loops. The motion of flare loops was fast in the early stage,
with an apparent velocity of about 110 km s$^{-1}$ (white dotted
line in panel (b)). Afterwards, the slippage became slow and the
slipping velocity decreased to 20$-$30 km s$^{-1}$ (blue dotted
lines in panel (b)). Here, the appearance of flare loops and their
slippage are intermittent, and indeed somewhat periodic, but each
time it is not the same ``clump" of brightness that is seen, but new
flare loops. About six strips were observed between 17:32 UT and
17:56 UT in the stack plot of 131 {\AA} (panel (b)). The time
intervals between two neighboring strips ranged from 3.3 min to 6.3
min, and the average period was about 4.8 min (panel (d)). These
bright features are seen to move for a period of several to 10 min
and their propagating distance were 10$-$22 Mm.

The quasi-periodic slipping motions are best seen at 94 and 131
{\AA} (panels (a)-(b)) and are not obvious in lower-temperature EUV
wavelengths (panel (c)). The 94 and 131 {\AA} channels respectively
correspond to high temperatures of about 7 MK and 11 MK, and the 171
{\AA} is at about 0.6 MK (O'Dwyer et al. 2010). In the stack plot of
171 {\AA}, the bright features are blurred and only four strips
could be discerned (panels (c) and (f)). The early peaks in each
light curve (panels (d) and (f)) appear at a similar place, however,
the late peaks are not consistent due to the decreasing intensity of
flare loops at 94 and 171 {\AA}. To study the evolution of moving
knots along the SNR, we construct the stack plot at 1400 {\AA}
(Figure 3(e)) along the straight line ``C$-$D" (Figure 2(d)). The
stack plot exhibits a similar quasi-periodic pattern, with several
approximately parallel strips of enhanced intensity. The moving
velocity of bright knots is similar to the slipping velocity of
flare loops. Seen from the 1400 {\AA} stack plot, the intersections
of the slipping knots along the SNR with the stationary slit of
\emph{IRIS} spectral observations could be clearly identified. The
six diamonds in the stack plot denote the time when the moving knots
arrived at the slit position.

To investigate the spectral characteristics of the moving knots
along the SNR, we use a Gaussian function to fit the profiles of Si
{\sc iv} 1402.77 {\AA} line and display the temporal evolution of
the Gaussian fit parameters in Figure 4. The temporal evolutions of
peak intensity, Doppler shift, and line width are in the slit range
between the two red horizontal bars in Figure 2(d). The six red
arrows in Figure 4 strictly correspond to the times of six diamonds
(Figure 3(e)) in the stack plot, when the slipping knots arrived at
the slit position. Thus we deduce that each wave-like structure in
Figure 4 is indicative of the process that a moving knot passes by
the slit position. A total of eight peaks could be discerned between
17:33 UT and 17:56 UT. The redshift exhibits a trend of an increase
at the locations of the moving knots. Before the appearance of the
knot in the slit, the redshift is about 10$-$20 km s$^{-1}$. This
value increases to about 30$-$50 km s$^{-1}$ near the peak time. The
line width of Si {\sc iv} at the location of the SNR shows a similar
trend. At the arrival of the slipping knot, the line profile becomes
broader and the width increases to 40$-$60 km s$^{-1}$.

\subsection{Spectroscopic Properties of the NNR and Post-flare Loops}

Similar to the SNR, the NNR also exhibits obvious redshifts in Si
{\sc iv} 1402.77 {\AA} line (Figure 5). In the impulsive phase of
the flare (between 17:21 and 17:45 UT), the NNR expanded in opposite
directions along the slit position and the ribbon became wider.
Along the slit, we select four locations: three corresponding to the
NNR and one at the background without flare ribbons (Figure 5(a)).
The profiles of the Si {\sc iv} line at these locations and their
Gaussian fittings are shown in Figure 5(e). For the spectral
profiles at some locations of the NNR are saturated in the impulsive
phase, the value of Doppler shift obtained by Gaussian fitting is
not exact and the uncertainty is estimated according to the 1-sigma
error in GAUSSFIT function. The redshift velocities at the outer
edges of the NNR are respectively 33$\pm$2 and 66$\pm$20 km s$^{-1}$
(red and yellow curves), larger than the velocity in the middle of
the ribbon (25 km s$^{-1}$; black curves). The temporal evolution of
the Doppler shift in the impulsive phase shows that the Si {\sc iv}
line is always redshifted at the NNR, with values of about 10$-$70
km s$^{-1}$ (panel (g)). By examining the spectroscopic observations
of other emission lines of Mg {\sc ii}, C {\sc ii} and Fe {\sc xii},
we find that these lines mainly exhibit obvious redshift at the NNR,
similar to the Si {\sc iv} line.

In the gradual phase of the flare (starting from 17:45 UT), the peak
intensities of the Si {\sc iv} line and redshift velocities both
exhibit a significant decrease (panels (e)-(f)). The maximum
intensity was at the southernmost edge of the NNR and the redshift
was only 11 km s$^{-1}$ at 17:56 UT (yellow curves in panel (f)).
The larger redshifts of 20$-$40 km s$^{-1}$ appeared at the north
part of the NNR (red and black curves in panel (f)). Starting from
17:52 UT, post-flare loops were formed as seen in SJI 1400 {\AA}
images (panel (b)). The emission of the Si {\sc iv} line at the
post-flare loops also shows an obvious redshift (blue curves in
panel (f)). The redshift velocities in the TR were about 80$-$100 km
s$^{-1}$ (panels (f)-(g)), within the range of the reconnection
outflow speed (Liu et al. 2013; Yang et al. 2015).

\section{Summary and Discussion}

We firstly report the observations of the quasi-periodic slipping
motion of flare loops during an X1.6 flare by the \emph{SDO} and
\emph{IRIS} on 2014 September 10. The slipping motion appeared in
one location along the SNR throughout the impulsive phase of the
flare. The associated period was 3$-$6 min and the speed of the
apparent slipping motion ranged from 20 to 110 km s$^{-1}$. The
footpoints of slipping loops were a series of bright knots that
simultaneously underwent a quasi-periodic slipping motion along the
flare ribbon. The spectroscopic observations of the \emph{IRIS}
showed that the redshifts of Si {\sc iv} 1402.77 {\AA} line at the
locations of moving knots were larger than other locations of the
flare ribbon, with a value of about 30$-$50 km s$^{-1}$. Similarly,
the line profile became broader at the locations of slipping knots
and the width increased to 40$-$60 km s$^{-1}$. For the NNR, the Si
{\sc iv} 1402.77 {\AA} line exhibited obvious redshifts of about
10$-$70 km s$^{-1}$ in the impulsive phase and the redshifts at the
outer edges of the ribbon were larger than those in the middle. The
post-flare loops were formed in the gradual phase of the flare and
the Si {\sc iv} line was redshifted by $\sim$ 80$-$100 km s$^{-1}$
near the loop footpoints.

The slipping is only an apparent motion since the connectivity of
magnetic field lines is continuously changing within the QSLs due to
a succession of reconnection between neighboring field lines
(Janvier et al. 2013; Dud{\'{\i}}k et al. 2014). The slipping motion
of flare loops shows a quasi-periodic pattern, indicating the
occurrence of quasi-periodic slipping reconnection. The slippage of
flare loops at the same location reappeared for several times, with
a time interval of 3$-$6 min. Dynamic models of magnetic
reconnection predict that the tearing mode instability and
coalescence of magnetic islands recur intermittently (Ofman et al.
1991; Kliem et al. 2000), resulting in the generation of oscillatory
reconnection. The triggering of magnetic reconnection and modulation
of its rate can be affected by fast MHD waves (Nakariakov \&
Melnikov 2009; McLaughlin et al. 2009), hence the magnetic
reconnection shows oscillatory dynamics. Heggland et al. (2009)
presented that the waves propagating upward from the
photosphere/convection zone induce periodic reconnection of the
magnetic field. Chen \& Priest (2006) proposed that p-mode
oscillations can induce oscillatory reconnection by modulating the
plasma density near the reconnection site. In our observations, the
period of the oscillatory reconnection is 3$-$6 min, which
approximately corresponds to the period of solar p-mode waves. One
possible explanation is that the p-mode oscillations above the
sunspots in AR 12158 may cause the variation of the plasma density
near the reconnection site and the reconnection rate oscillates
accordingly (Chen \& Priest 2006). Thus the slipping magnetic
reconnection appears in a quasi-periodic way and the slipping motion
of flare loops is intermittent and quasi-periodic.

In the flare ribbons, the \emph{IRIS} emission lines (Mg {\sc ii}, C
{\sc ii}, Si {\sc iv} and Fe {\sc xii}) all exhibited significant
redshifts (the Si {\sc iv} line for 10$-$70 km s$^{-1}$). Redshifts
measured in plasma of the TR are usual in plages (Klimchuk 1989;
Simon et al. 1986; Malherbe et al. 1987). The velocity measured in
the middle of bright ribbons is blueshifted (Schmieder et al. 1987;
Berlicki et al. 2005). In the impulsive phase of the flare, the hot
emission lines at temperatures of several MK mostly exhibited
blueshifts with speeds of several hundred km s$^{-1}$ while cool
emission lines were redshifted by tens of km s$^{-1}$ (del Zanna et
al. 2006; Li \& Ding 2011; Ning 2011; Tian et al. 2014b). In our
work, the redshifted ribbons in the TR probably imply the
downward-moving cool and dense chromospheric and TR condensations
(Fisher et al. 1985) during the impulsive phase and the early stage
of the gradual phase.

\acknowledgments {Many thanks to Hardi Peter, Hui Tian, Le-Ping Li
and Li-Mei Yan for discussion about data analysis. \emph{IRIS} is a
NASA small explorer mission developed and operated by LMSAL with
mission operations executed at NASA Ames Research center and major
contributions to downlink communications funded by the Norwegian
Space Center (NSC, Norway) through an ESA PRODEX contract. We
acknowledge the \emph{SDO}/AIA and HMI for providing data. This work
is supported by the National Basic Research Program of China under
grant 2011CB811403, the National Natural Science Foundations of
China (11303050, 11025315, 11221063 and 11003026), the CAS Project
KJCX2-EW-T07 and the Strategic Priority Research Program$-$The
Emergence of Cosmological Structures of the Chinese Academy of
Sciences, Grant No. XDB09000000.}

{}
\clearpage

\begin{figure}
\centering
\includegraphics
[bb=14 209 547 611,clip,angle=0,scale=0.85]{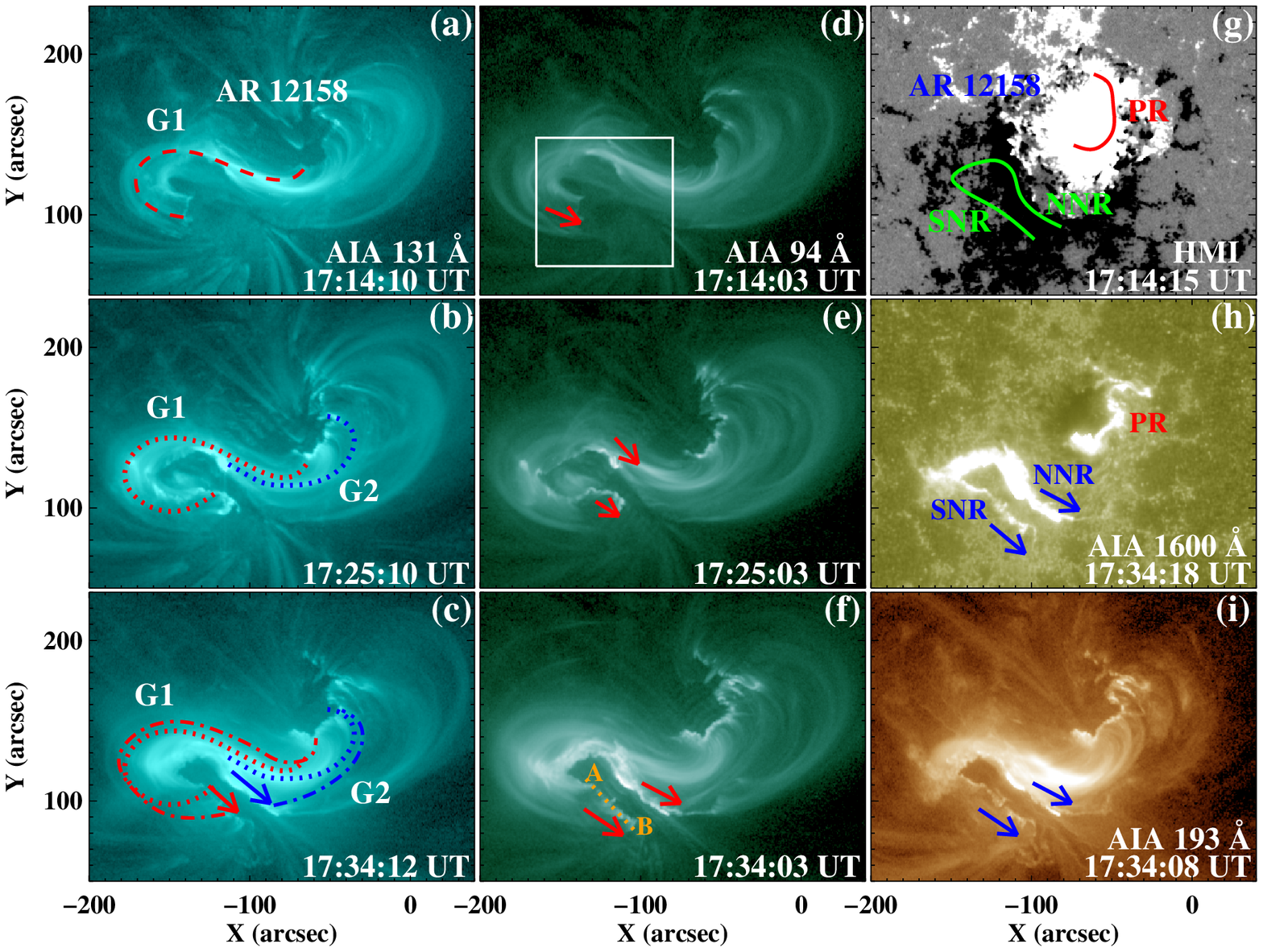}
\caption{Evolution of flare loops and flare ribbons (PR, NNR and
SNR) during the X1.6 flare on 2014 September 10 (see Animation
131-slippage). ``G1" and ``G2" in panels (a)-(c) represent two
groups of slipping flare loops. Dotted lines in panel (c) are the
duplicates of the flare loops in panel (b). The arrows point to the
propagating directions of NNR and SNR. The white rectangle in panel
(d) denotes the FOV of Figures 2(a)-(d). Dotted line ``A$-$B" (panel
(f)) shows the cut position used to obtain the stack plots shown in
Figures 3(a)-(c). Red and green curves in panel (g) denote the
ribbons PR, NNR and SNR. The arrows in panels (h) and (i) show the
propagating directions of NNR and SNR. \label{fig1}}
\end{figure}
\clearpage

\begin{figure}
\centering
\includegraphics
[bb=9 273 554 544,clip,angle=0,scale=0.85]{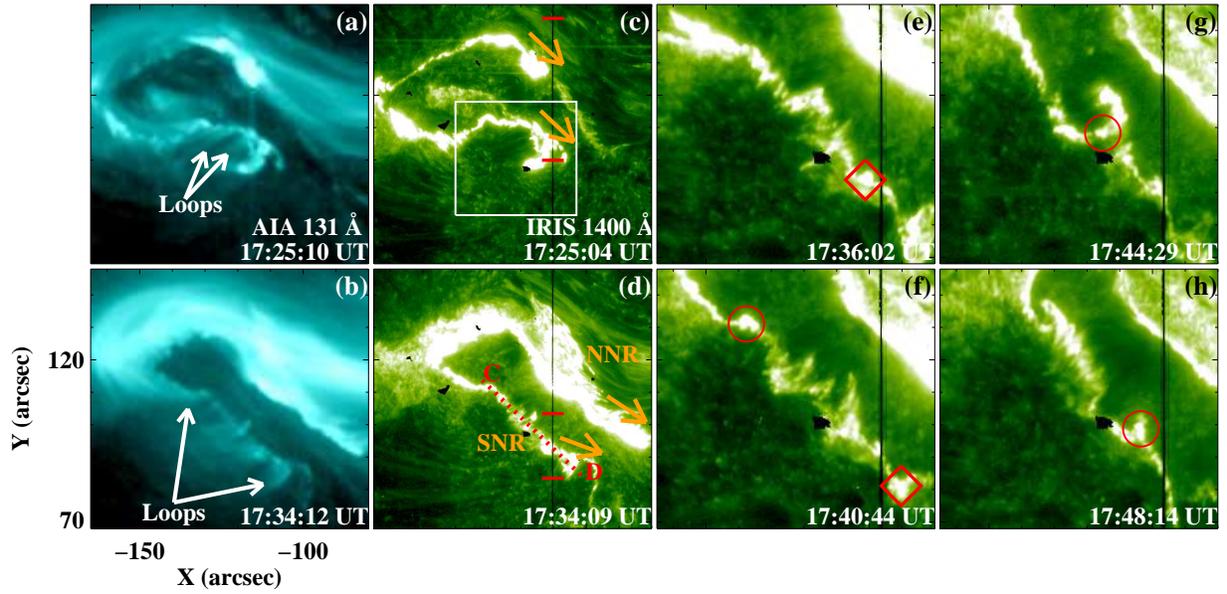} \caption{AIA
131 {\AA} (panels (a)-(b)) and \emph{IRIS} 1400 {\AA} images (panels
(c)-(h)) showing the slipping loops and fine structures of the flare
ribbon (see Animation 1400-ribbon). White rectangle in panel (c)
denotes the FOV of panels (e)-(h). Dotted line ``C$-$D" (panel (d))
shows the cut position used to obtain the stack plot shown in Figure
3(e). The two red horizontal bars in panels (d) and (c) respectively
indicate the spatial range shown in Figures 4 and 5. The diamonds
and circles (panels (e)-(h)) outline two propagating bright knots
along the SNR. \label{fig2}}
\end{figure}
\clearpage

\begin{figure}
\centering
\includegraphics
[bb=22 140 571 678,clip,angle=0,scale=0.8]{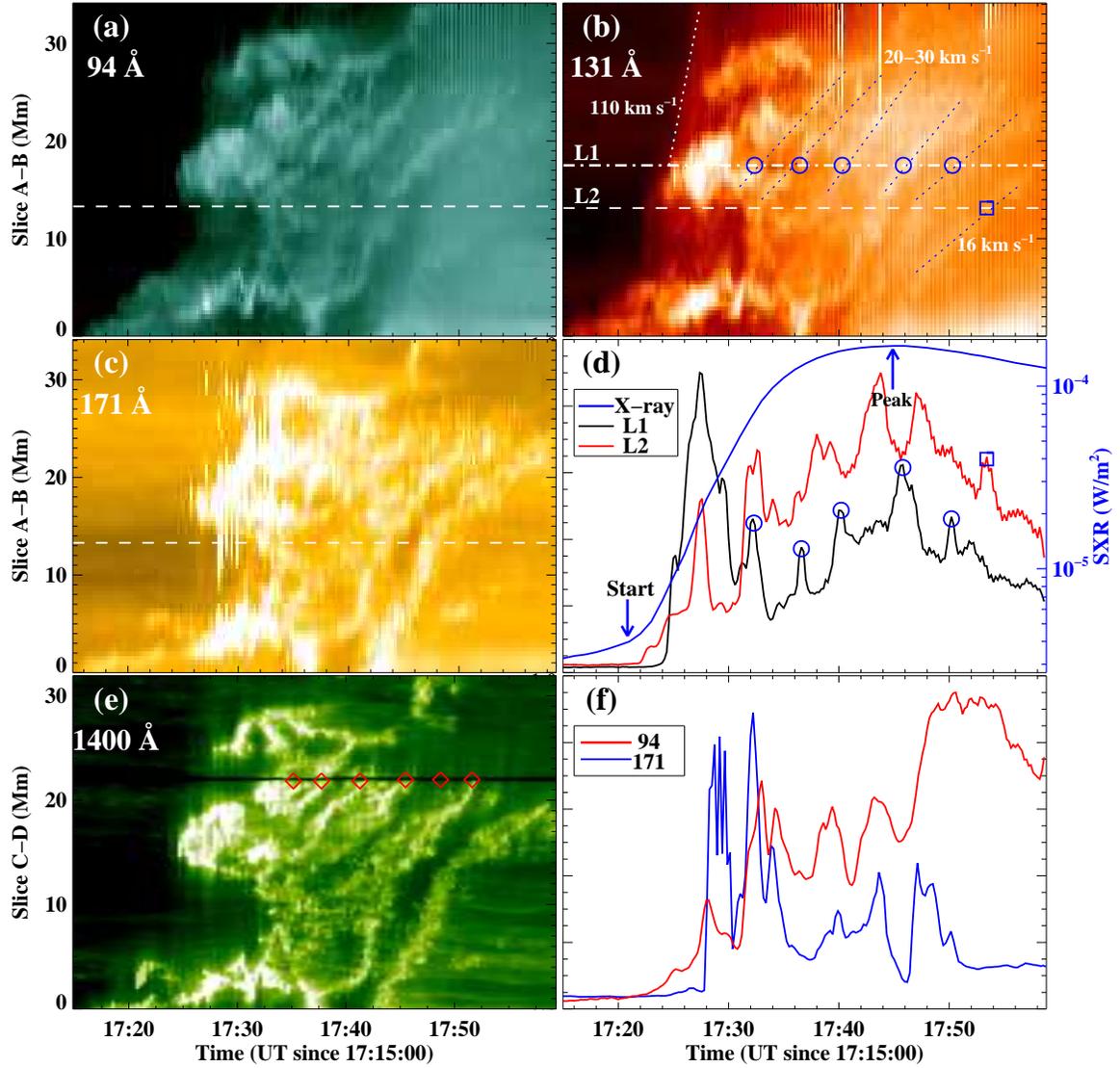} \caption{Panels
(a)-(c): stack plots along slice ``A$-$B" (yellow dotted line in
Figure 1(f)) in different wavelengths showing the quasi-periodic
slipping motion. Blue circles and square in panel (b) denote the
intersections of blue dotted lines with two horizontal lines (``L1"
and ``L2"). Panel (d): GOES SXR 1$-$8 {\AA} flux (blue curve) of the
associated flare and horizontal slices (black and red curves) along
the dash-dotted (``L1") and dashed lines (``L2") in panel (b). Blue
circles and square are the duplicates of symbols in panel (b). Panel
(e): stack plot along slice ``C$-$D" (red dotted line in Figure
2(d)) at 1400 {\AA} showing the evolution of the SNR. The six
diamonds denote the time when the moving knots arrive at the slit
position. Panel (f): horizontal slices along the dashed lines in
stack plots of 94 and 171 {\AA}. \label{fig3}}
\end{figure}
\clearpage

\begin{figure}
\centering
\includegraphics
[bb=110 161 484 662,clip,angle=0,scale=0.9]{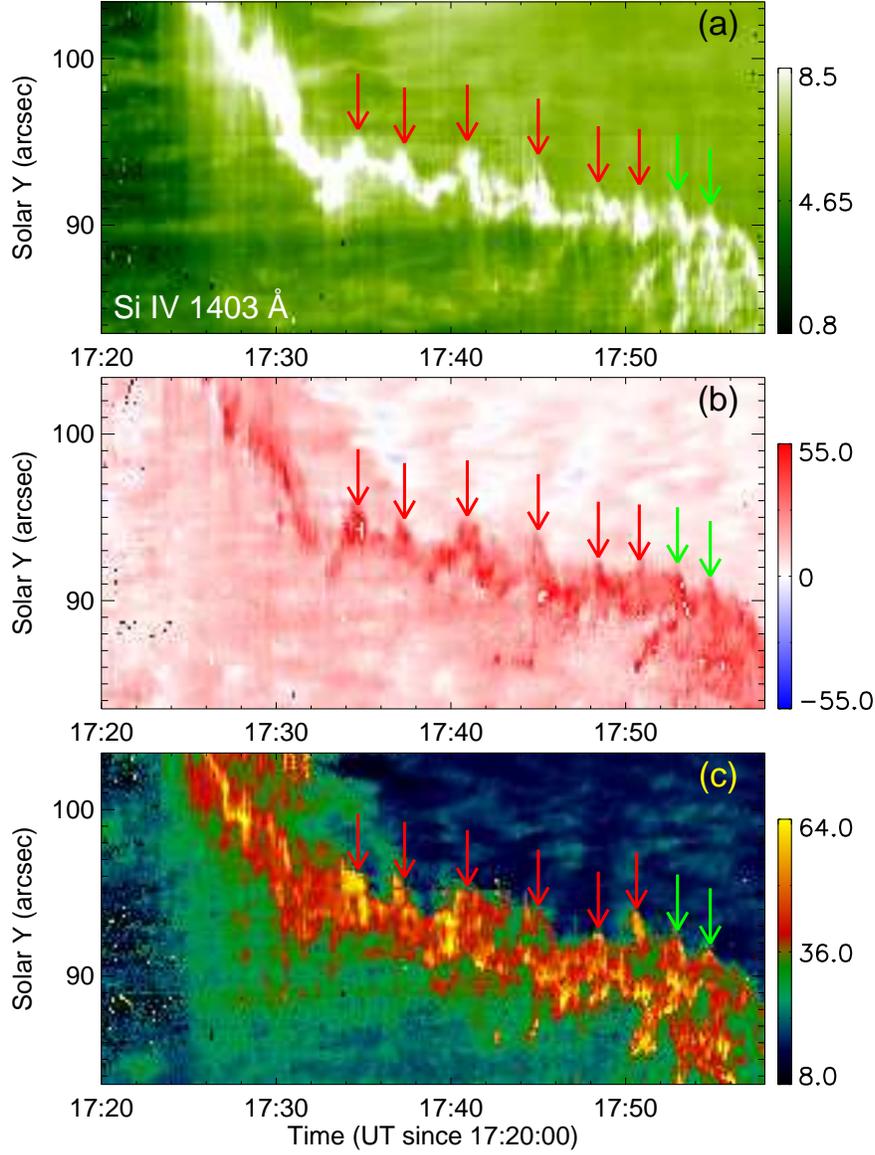}
\caption{Temporal evolution of peak intensity (a), Doppler shift
(b), and line width (c) in the spatial range indicated by the two
red horizontal bars in Figure 2(d). These parameters are obtained by
applying single-Gaussian fits to the Si {\sc iv} 1402.77 {\AA} line.
The arrows point to the peaks of wave-like evolution and correspond
to the times when the slipping knots pass by the location of the
\emph{IRIS} slit (the six red arrows correspond to the six diamonds
in Figure 3(e); two green arrows are not observed in the stack plot
of 1400 {\AA} for the propagating direction of the SNR is away from
the slice ``C$-$D"). \label{fig4}}
\end{figure}
\clearpage

\begin{figure}
\centering
\includegraphics
[bb=53 157 549 673,clip,angle=0,scale=0.9]{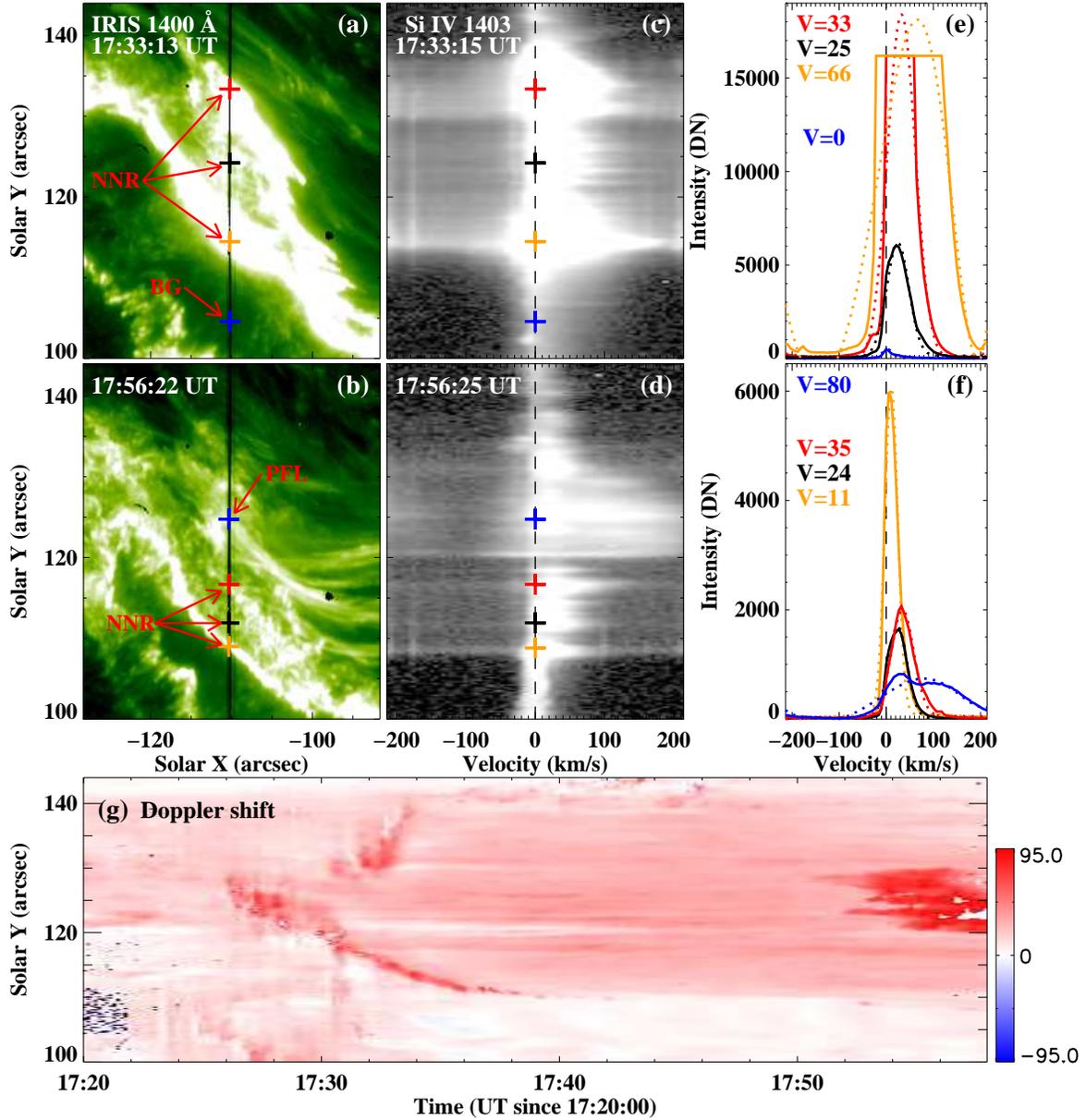} \caption{Panels
(a)-(b): \emph{IRIS} 1400 {\AA} images showing the NNR and the
post-flare loops (PFL). The range of y axis is between the two red
horizontal bars in Figure 2(c). The red, black and yellow pluses in
panels (a) and (b) denote the three locations in the NNR. The blue
plus (BG) in panel (a) indicates one location at the background
without flare ribbons, and the blue one (PFL) in panel (b) displays
the intersection of the PFL with the slit. Panels (c)-(d): evolution
of the Si {\sc iv} 1402.77 {\AA} spectra in the slit range of panels
(a)-(b). Panels (e)-(f): profiles of the Si {\sc iv} line at the
NNR, BG and PFL. The dotted curves are the Gaussian fitting
profiles. Panel (g): temporal evolution of the Si {\sc iv} Doppler
shift at the NNR and PFL. The spatial range is the same as that of
panels (a)-(d). \label{fig5}}
\end{figure}
\clearpage

\end{document}